\documentclass[epj]{webofc}
\usepackage[varg]{txfonts}   % Web of Conferences font

\usepackage{amsmath,amssymb,amsfonts,graphicx}
\usepackage{bm}
\usepackage{slashed}
\usepackage{subfigure}
\usepackage[svgnames]{xcolor}
\newcommand{\vect}[1]{\boldsymbol{#1}}

\newcommand{\al}[1]{\begin{align} #1 \end{align}}

\newcommand{\Ge}{\mathrm{GeV}}
\newcommand{\Gs}{\mathrm{GeV}^2}
\wocname{EPJ Web of Conferences}
\woctitle{INPC 2013}
\begin{document}
\title{The Effect of Vector Meson Decays on Dihadron Fragmentation Functions}
\author{Hrayr~H.~Matevosyan\inst{1}\fnsep\thanks{\email{hrayr.matevosyan@adelaide.edu.au}} \and
       Anthony~W.~Thomas\inst{1}         % etc.
       \and Wolfgang Bentz\inst{2}
}
\institute{CSSM and ARC Centre of Excellence for Particle Physics at the Tera-scale,\\ 
School of Chemistry and Physics, \\
University of Adelaide, Adelaide SA 5005, Australia
\\ http://www.physics.adelaide.edu.au/cssm 
\and
Department of Physics, School of Science,\\  Tokai University, Hiratsuka-shi, Kanagawa 259-1292, Japan
\\ http://www.sp.u-tokai.ac.jp/
}
\abstract{%
Dihadron Fragmentation Functions (DFF) provide a vast amount of information on the intricate details of the parton hadronization process. Moreover, they provide a unique access to the "clean" extraction of nucleon transversity parton distribution functions in semi inclusive deep inelastic two hadron production process with a transversely polarised target. On the example of the $u\to\pi^+\pi^-$, we analyse the properties of unpolarised DFFs using their probabilistic interpretation. We use both the NJL-jet hadronization model and PYTHIA 8.1 event generator to explore the effect of the  strong decays  of the vector mesons produced in the quark hadronization process on the pseudoscalar DFFs. Our study shows that, even though it is less probable to produce vector mesons in the hadronization process than pseudo scalar mesons of the same charge, the products of their strong decays drastically affect the DFFs for pions because of the large combinatorial factors. Thus, an accurate description of both vector meson production and decays are crucial for theoretical understanding of DFFs.
}
\maketitle
%
%%%%%%%%%%%%%%%%%%%%%%%%%%%%%%%%%%%%%%%%%%%%%%%%%%%%%
%%%%%%%%%%%%%%%%%%%%%%%%%%%%%%%%%%%%%%%%%%%%%%%%%%%%%
%%%%%%%%%%%%%%%%%%%%%   SECTION %%%%%%%%%%%%%%%%%%%%%%%%%%
\section{Introduction}
\label{intro}

 The parton hadronization process is one of the most challenging aspects of the theoretical description of strongly interacting systems, as it describes the transition of the fast moving partons from perturbative regime into non-perturbative hadronic bound states. The probability of production of a single hadron in this process is described by the ordinary fragmentation functions (FF), which have been studied in both theoretical models and  extracted from the experimental data using various parametrisation procedures. Nevertheless, there is still a large uncertainty in our knowledge of FFs, especially in the case of the unfavored and kaon channels. This has been demonstrated in the recent experimental results from SIDIS measurements ~\cite{Airapetian:2012ki, Makke:2013bya, Makke:2013dya}, where the existing parametrizations of the fragmentation and parton distribution functions failed to describe the measured multiplicities. 
 
 The dihadron fragmentation functions (DFF) describe the production of two hadrons in parton hadronization process, which brings an extra layer of complexity for both theoretical description and experimental extraction. For theoretical models of DFF, a detailed picture of the hadronization final states should be given, as the leading hadron approximation typically used in FF models is insufficient here. On the other hand, the comparison of the DFFs with experimental extractions would serve as an extra constraint on these models which provide a complete hadronization description, such as the Lund model~\cite{Sjostrand:1982fn} implemented in the PYTHIA event generator~\cite{Sjostrand:2006za,Sjostrand:2007gs} and those based on the quark-jet model~\cite{Field:1976ve,Field:1977fa,Casey:2012ux,Casey:2012hg}. 
 
 The importance of exploring DFFs extends beyond the description of the hadronization mechanism. It was realised~\cite{Radici:2001na} that in semi-inclusive deep inelastic scattering (SIDIS) with two recorded final hadrons, the corresponding cross section can be factorized into nucleon parton distribution function (PDF), a hard scattering cross-section and a DFF. More importantly, instead of the transverse momentum convolution of PDF and FF in a single hadron SIDIS, here when integrating over the transverse moment of both hadrons we obtain a simple product of collinear PDF and a DFF that depends on the sum of the produced hadrons' light-cone momentum fractions and their invariant mass. Such a separation gives direct access to the nucleon transversity PDF through a measurement of a single spin asymmetry in SIDIS with transversely polarised target, though a knowledge of the corresponding DFFs is required. The first such extraction was performed in Ref.~\cite{Bacchetta:2011ip} using the SIDIS two hadron asymmetry measured by HERMES~\cite{Airapetian:2008sk} and convolution of DFFs in two hadron pair production in $e^+e^-$ annihilation in BELLE~\cite{Vossen:2011fk}. Unfortunately, we have limited information on both unpolarised and interference DFFs, thus a number of assumptions and extrapolations had to be employed in this extraction. Moreover, the most recent model for DFFs  was constructed in Ref.~\cite{Bacchetta:2006un}, where the model parameters were fixed by fitting the unpolarised DFFs for $\pi^+\pi^-$ pairs to Monte Carlo (MC) sample generated using PYTHIA framework. 
 
 Our ultimate goal is to model the DFFs in the NJL-jet model~\cite{Ito:2009zc,Matevosyan:2010hh,Matevosyan:2011ey,PhysRevD.86.059904,Matevosyan:2011vj,Matevosyan:2012ga,Matevosyan:2012ms}, where a complete quark hadronization description is given. This work is underway and will be reported in our upcoming paper~\cite{Matevosyan:2013aa}. In this work we explore one important aspect of the hadronization process that strongly affects DFFs, namely how the production and strong decays of vector mesons during quark hadronization affect the observed pion unpolarised DFFs. Naively following our earlier work involving a similar study in the case of FFs, that showed a modest contribution of vector meson decay products with respect to directly produced hadrons~\cite{Matevosyan:2011ey,PhysRevD.86.059904}, one expects a similar picture to hold for DFFs. 
 
  In the next Section we briefly introduce the NJL-jet model and our method of calculating the DFFs using MC method. In Section~\ref{SEC_RES_NJL}  we will present our results for the NJL-jet model calculations of unpolarised DFFs. In Section~\ref{SEC_RES_PYTHIA}, we will present analogous results obtained using PYTHIA framework to test the model dependence of our results, which will be followed by Section~\ref{SEC_CONC} with conclusions and outlook.
  
%%%%%%%%%%%%%%%%%%%%%%%%%%%%%%%%%%%%%%%%%%%%%%%%%%%%%
%%%%%%%%%%%%%%%%%%%%%%%%%%%%%%%%%%%%%%%%%%%%%%%%%%%%%
%%%%%%%%%%%%%%%%%%%%%   SECTION %%%%%%%%%%%%%%%%%%%%%%%%%%
\section{Calculating DFFs in the NJL-jet Model}
\label{SEC_CALC_DFF}

%%%%%%%%%%%%%%%%%%%%%%%%%%%%%%%%%%%%%%%%%%%%%%%%%%%%%
%%%%%%%%%%%%%%%%%%%%%%%%%%%%%%%%%%%%%%%%%%%%%%%%%%%%%
%%%%%%%%%%%%%%%%%%%%%   SUB-SECTION %%%%%%%%%%%%%%%%%%%%%%%%%%
\subsection{The NJL-jet Model}
\label{SUB_SEC_NJL}

\begin{figure}[h]
\centering
\includegraphics[width=10cm,clip]{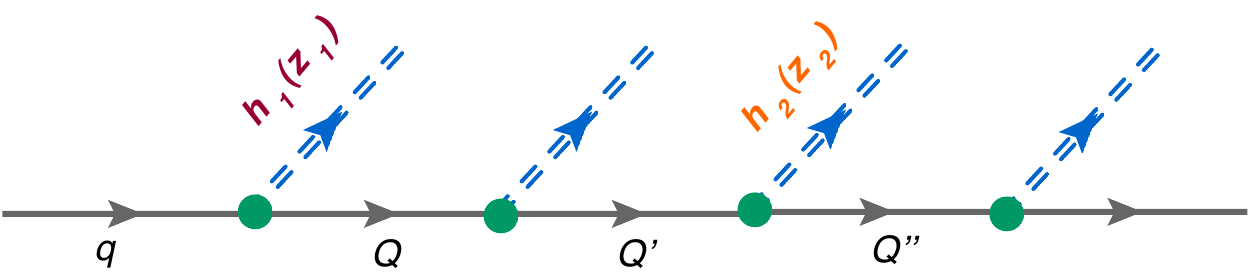}
\caption{The quark-jet hadronization mechanism.}
\label{PLOT_QUARK_JET}
\end{figure}
 
 The NJL-jet model~\cite{Ito:2009zc,Matevosyan:2010hh,Matevosyan:2011ey,PhysRevD.86.059904,Matevosyan:2011vj,Matevosyan:2012ga,Matevosyan:2012ms} uses the Field and Feynman's original quark-jet hadronization framework~\cite{Field:1976ve,Field:1977fa}, where the fragmenting quark sequentially emits hadrons that do not re-interact, as schematically depicted in Fig.~\ref{PLOT_QUARK_JET}. The elementary hadron emission probabilities at each vertex are calculated using the effective quark model of Nambu and Jona-Lasinio (NJL)~\cite{Nambu:1961tp,Nambu:1961fr}. The fragmentation functions are extracted from the corresponding hadron multiplicities, calculated as Monte Carlo averages of the  hadronization process of a quark with a given flavour when restricting the total number of emitted hadrons for each fragmentation chain to a predefined number. In the original formalism of Field and Feynman, the number of the produced hadrons were taken to  be infinite, yielding a coupled set of integral equations for both FFs and DFFs.
 
  The first study of DFFs that depend on the light-cone momentum fractions of each produced hadron within the NJL-jet model was performed in Ref.~\cite{Casey:2012ux}. The DFFs for pion, kaon and mixed pairs were calculated  using the integral equation formalism, while their scale evolution was studied in Ref.~\cite{Casey:2012hg}. In these papers, only pion and kaon emission channels were considered.
  
  In the current work we consider the hadronization of both light and strange quarks to $\pi$, $K$, $\rho$, $\omega$, $K^*$ and $\phi$ mesons, along with the two- and three- body strong decays of the vector mesons (as fragmentation functions encode only the strong interaction physics.) Throughout this article we only present the results for $u\to\pi^+\pi^-$. The details of the calculations, including the elementary fragmentation functions used in all the channels and the vector meson decay cross-section calculations in light-cone variables, as well as the results for the many possible hadron pairs will be given in our upcoming publication~\cite{Matevosyan:2013aa}.
 
%%%%%%%%%%%%%%%%%%%%%%%%%%%%%%%%%%%%%%%%%%%%%%%%%%%%%
%%%%%%%%%%%%%%%%%%%%%%%%%%%%%%%%%%%%%%%%%%%%%%%%%%%%%
%%%%%%%%%%%%%%%%%%%%%   SUB-SECTION %%%%%%%%%%%%%%%%%%%%%%%%%%
\subsection{Calculating DFFs using MC}
\label{SUB_SEC_MC}

 Here we consider the fragmentation of a quark of flavour $q$ to two hadrons $h_1$ and $h_2$, following the standard kinematic conventions (see Ref.~\cite{Radici:2001na}). Let's choose the coordinate system such that the $z$ axis is along the direction of the 3-momentum of the initial fragmenting quark $q$. Then the momenta of the quark and the two produced hadrons of interest $h_1$ and $h_2$ can be represented as
\begin{align}
\label{EQ_MOMENTA}
&P_q=(k^-,k^+,\vect{0}),\\ \nonumber
&P_{h1}\equiv P_1 = (z_1 k^-, P_1^+, \vect{P}_{1\perp}),\ P_1^2 = M_{h1}^2, \\ \nonumber
&P_{h2}\equiv P_2 = (z_2 k^-, P_2^+, \vect{P}_{2\perp}),\ P_2^2=M_{h2}^2,
\end{align}
where $z_1, M_{h1}$ and $z_2, M_{h2}$ are the corresponding light-cone momentum fractions and the masses of the hadrons. 

Here we want to calculate the dihadron fragmentation functions $D_q^{h_1h_2}(z,M_h^2)$ as a function of the sum of the light-cone momentum fractions $z=z_1+z_2$ and the invariant mass square $M_h^2=(P_1+P_2)^2$ of the produced hadron pair. We employ the number density interpretation for the $D_q^{h_1h_2}(z,M_h^2)$ to extract them by calculating the corresponding multiplicities using MC average over simulations of the quark hadronization process, similar to the method employed for the single hadron FF extractions~\cite{Matevosyan:2011ey,PhysRevD.86.059904,Matevosyan:2011vj,Matevosyan:2012ga}
\begin{align}
\label{EQ_MC_EXTRACT}
D_q^{h_1h_2}(z,M_h^2)\ \Delta z\ \Delta M_h^2
= \left< N_{q}^{h_1 h_2}(z,z+\Delta z; M_h^2,M_h^2  +\Delta M_h^2 )\right>.
\end{align}

 For each MC simulation, we consider all the directly produced (so-called "primary") hadrons by the quark $q$, and calculate the $z$ and $ M_h^2$ for all the hadron pairs, filling-in the corresponding histograms. We also repeat the procedure by now considering all the final state hadrons after allowing for the strong decays of the "primary" produced resonances, particularly the vector mesons.

%%%%%%%%%%%%%%%%%%%%%%%%%%%%%%%%%%%%%%%%%%%%%%%%%%%%%
%%%%%%%%%%%%%%%%%%%%%%%%%%%%%%%%%%%%%%%%%%%%%%%%%%%%%
%%%%%%%%%%%%%%%%%%%%%   SECTION %%%%%%%%%%%%%%%%%%%%%%%%%%
\section{Results from the NJL-jet Model}
\label{SEC_RES_NJL}

 We performed MC simulations using our NJL-jet model based software framework and calculated the relevant DFFs using the formula in Eq.~(\ref{EQ_MC_EXTRACT}). In this calculation, we performed $10^{10}$ simulations and used $500$ bins to discretize the values of $z$ in the region $[0;1]$, as well as $25000$ bins for $M_h^2$  in the range $[0;5]~\mathrm{GeV}^2$. 
 
 %===============================================================================
\begin{figure}[htb]
\centering 
\subfigure[] {
\includegraphics[width=0.45\columnwidth]{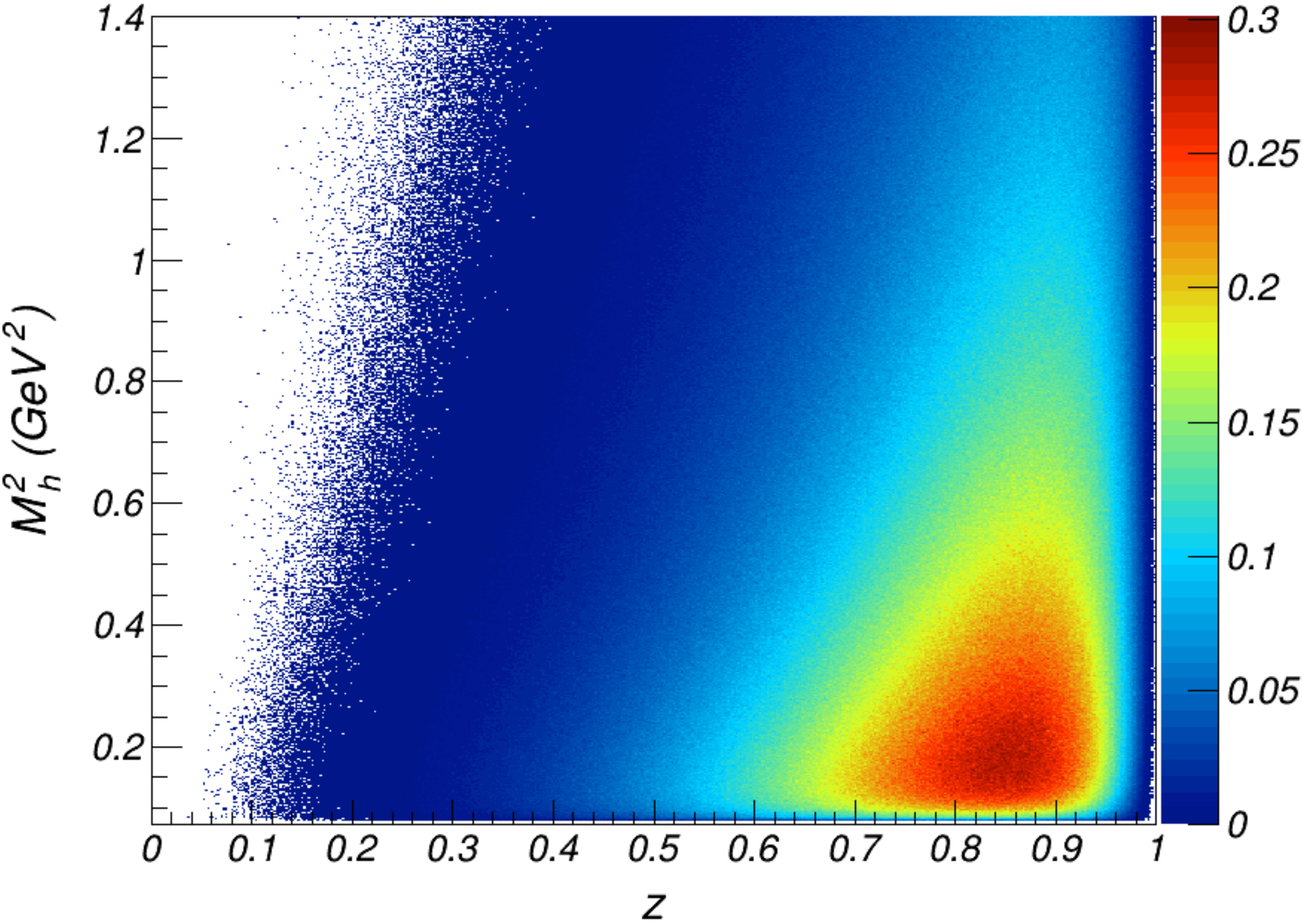}
}
\hspace{0.1cm} 
\subfigure[] {
\includegraphics[width=0.45\columnwidth]{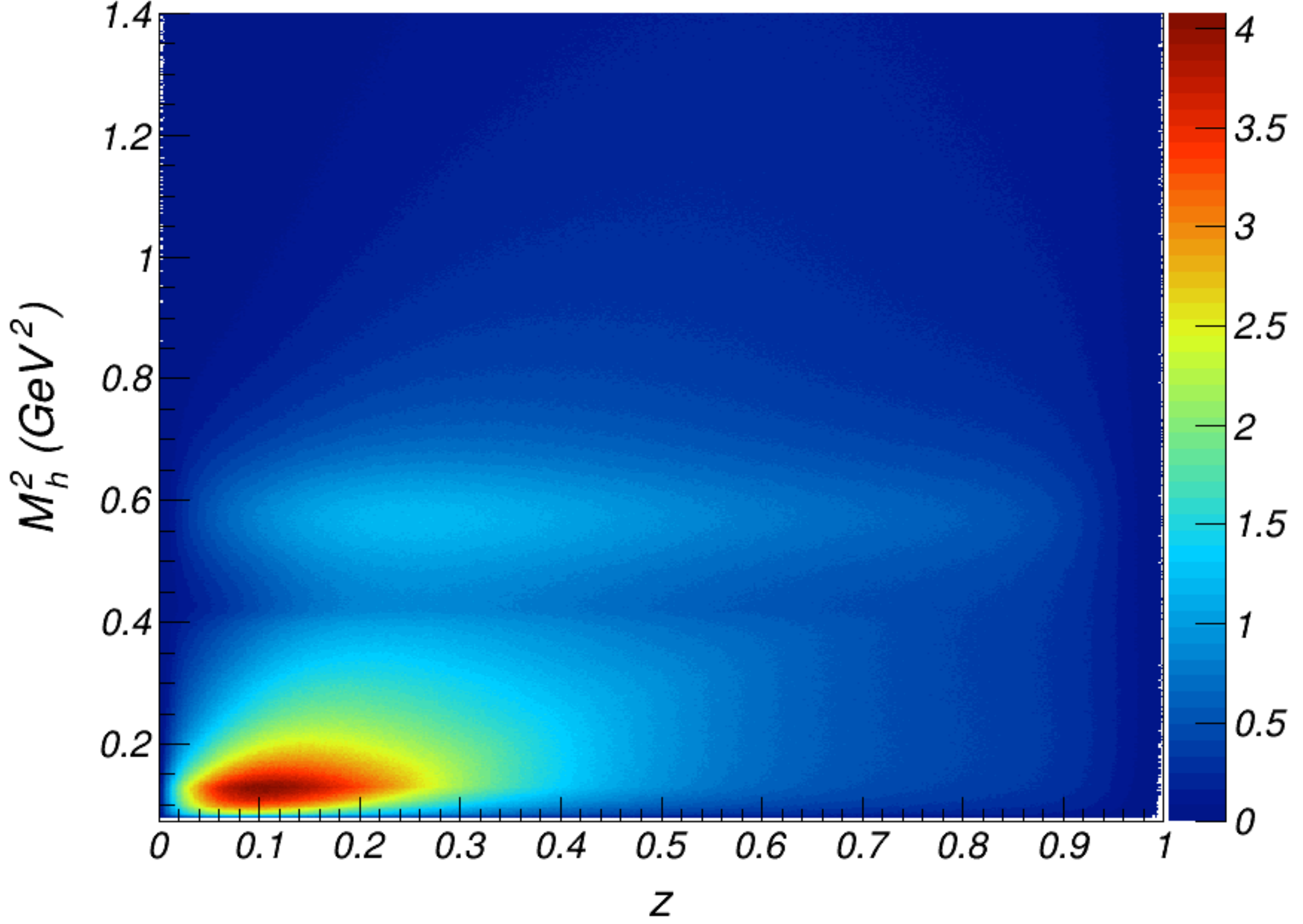}
}
\caption{The results for $D_u^{\pi^+\pi^-}$ for (a) only primary and (b) all the final state hadrons after decay of vector mesons from the NJL-jet MC simulations with $2$ primary produced hadrons.}
\label{PLOT_DFF_3D}
\end{figure}
%===============================================================================

\begin{figure}[htb]
\centering 
\subfigure[] {
\includegraphics[width=0.45\columnwidth]{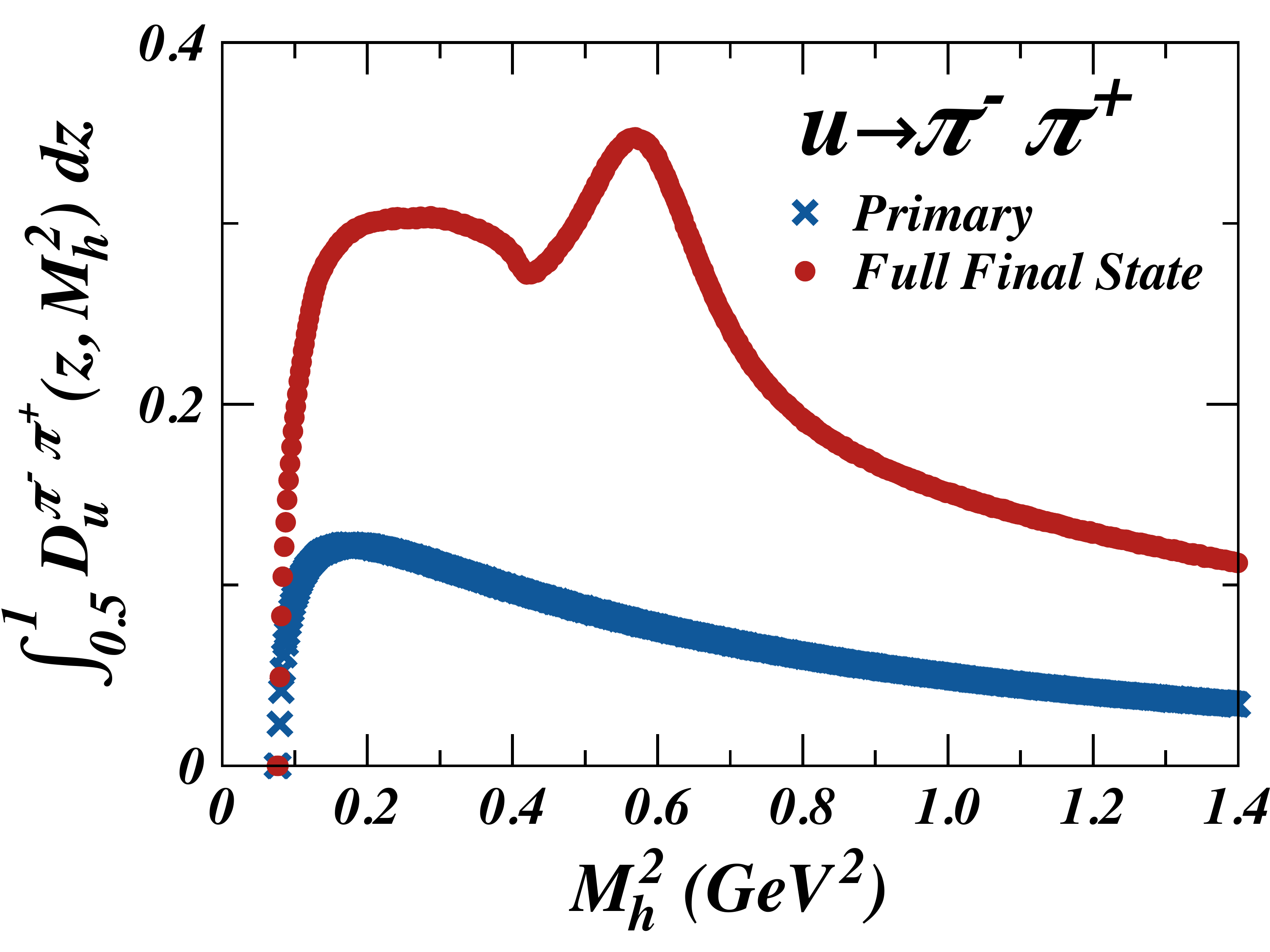}
}
\hspace{0.1cm} 
\subfigure[] {
\includegraphics[width=0.45\columnwidth]{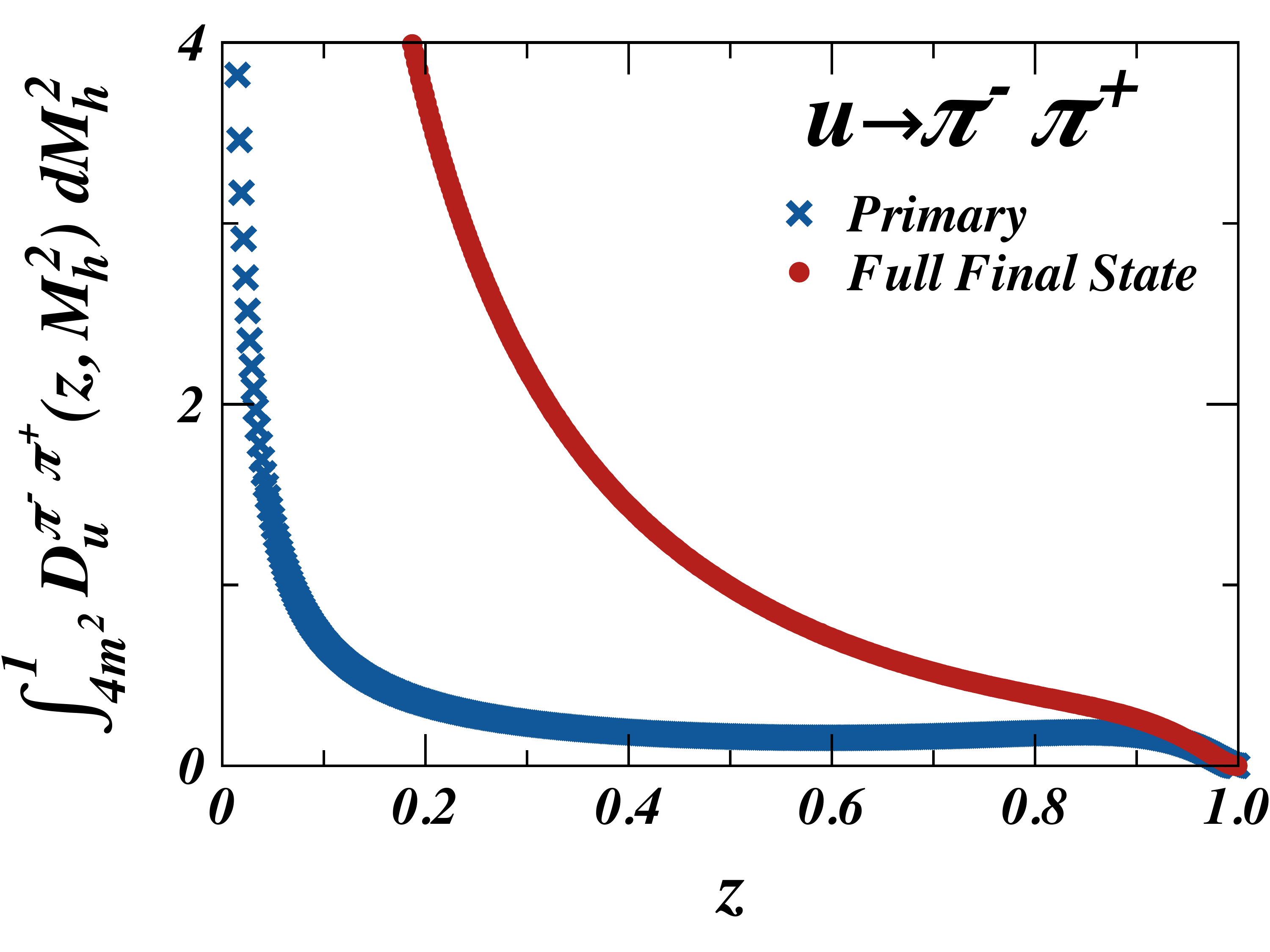}
}
\caption{The comparison of the results for $D_u^{\pi^+\pi^-}$ for the primary (blue crosses) and full (red dots) final state when (a) integrated over $z$ and (b) integrated over $M_h^2$. The MC simulations were performed with $8$ primary produced hadrons and over $10^{10}$ simulations.}
\label{PLOT_DFF_INT}
\end{figure}

  The results for the $D_u^{\pi^+\pi^-}(z,M_h^2)$ extracted from simulations with $2$ produced primary hadrons are shown on the plots in Fig.~\ref{PLOT_DFF_3D}. Here the plot in Fig.~\ref{PLOT_DFF_3D}(a) depicts the results obtained when considering only the pions produced directly by the fragmenting $u$ quark (primary pions), while Fig.~\ref{PLOT_DFF_3D}(b) depicts the results when also considering the pions that are produced in the decays of the primary vector mesons (secondary hadrons). We see a dramatic effect when considering the full final state (FFS), unexpected from the naive analogy to the moderate contribution of the secondary hadrons  to FFs (see Refs.~\cite{Matevosyan:2011ey,PhysRevD.86.059904}). We also note the distinct signature of the $\rho^0$ meson peak around $M_h^2\approx 0.77^2 \Gs $, as well as the enhancement in the small $z$ and invariant mass region due to $\omega$ meson decay. To investigate in detail these features it is useful to integrate the DFF over some range of either $z$ or $M_h^2$. The plots in Fig.~\ref{PLOT_DFF_INT} depict the comparison of the results with primary and full set of final hadrons for $D_u^{\pi^+\pi^-}$, integrated over (a) $z$ in the region $0.5$ to $1$ and (b) over $M_h^2$ in the region $4m_\pi^2\approx 0.08~\Gs$ to $1~\Gs$. We readily see on the plots in Fig.~\ref{PLOT_DFF_INT}(a), depicting the invariant mass squared dependence of the DFF, the pronounced  effect of the vector meson resonances, such as the $\rho^0$ peak and the enhancement in the region below $0.4~\Gs$ coming from the $\omega\to \pi^+\pi^-\pi^0$ decay with shifted invariant mass due to unobserved $\pi^0$. 
  
%===============================================================================
%===============================================================================

%%%%%%%%%%%%%%%%%%%%%%%%%%%%%%%%%%%%%%%%%%%%%%%%%%%%%
%%%%%%%%%%%%%%%%%%%%%%%%%%%%%%%%%%%%%%%%%%%%%%%%%%%%%
%%%%%%%%%%%%%%%%%%%%%   SECTION %%%%%%%%%%%%%%%%%%%%%%%%%%
\section{DFF Studies using PYTHIA}
\label{SEC_RES_PYTHIA}

 The stark effect of the vector meson decays on the DFFs has prompted us to test the model dependence of our findings. We have employed PYTHIA 8.1 event generator \cite{Sjostrand:2007gs} to perform analogous extractions of DFFs from MC average of corresponding multiplicities using the formula in Eq.~(\ref{EQ_MC_EXTRACT}). To accomplish this, we need to find all the hadrons that are produced in the hadronization of the given initial quark, which is not straight-forward in the Lund string fragmentation model employed in PYTHIA. Our method is to setup an initial $q\bar{q}$ hard system with momenta along the $z$ axis (positive for $q$ and negative for $\bar{q}$),  and allow only hadronization. Then all the produced hadrons with positive $z$-component of 3-momentum are assigned to $q$ and the ones with negative $z$ momentum to $\bar{q}$. We also allow only the same strong decay channels of the produced vector mesons that we used in the NJL-jet model. Here we can determine the primary and secondary hadrons using the information provided by the generator. Further, we extract analogously the DFFs with primary and FFS hadrons by calculating the MC average of the relevant multiplicities, repeating the hadronization of $q\bar{q}$ system with the same flavour and centre-of-mass energy to achieve sufficient statistical precision. 
 
\begin{figure}[htb]
\centering 
\subfigure[] {
\includegraphics[width=0.45\columnwidth]{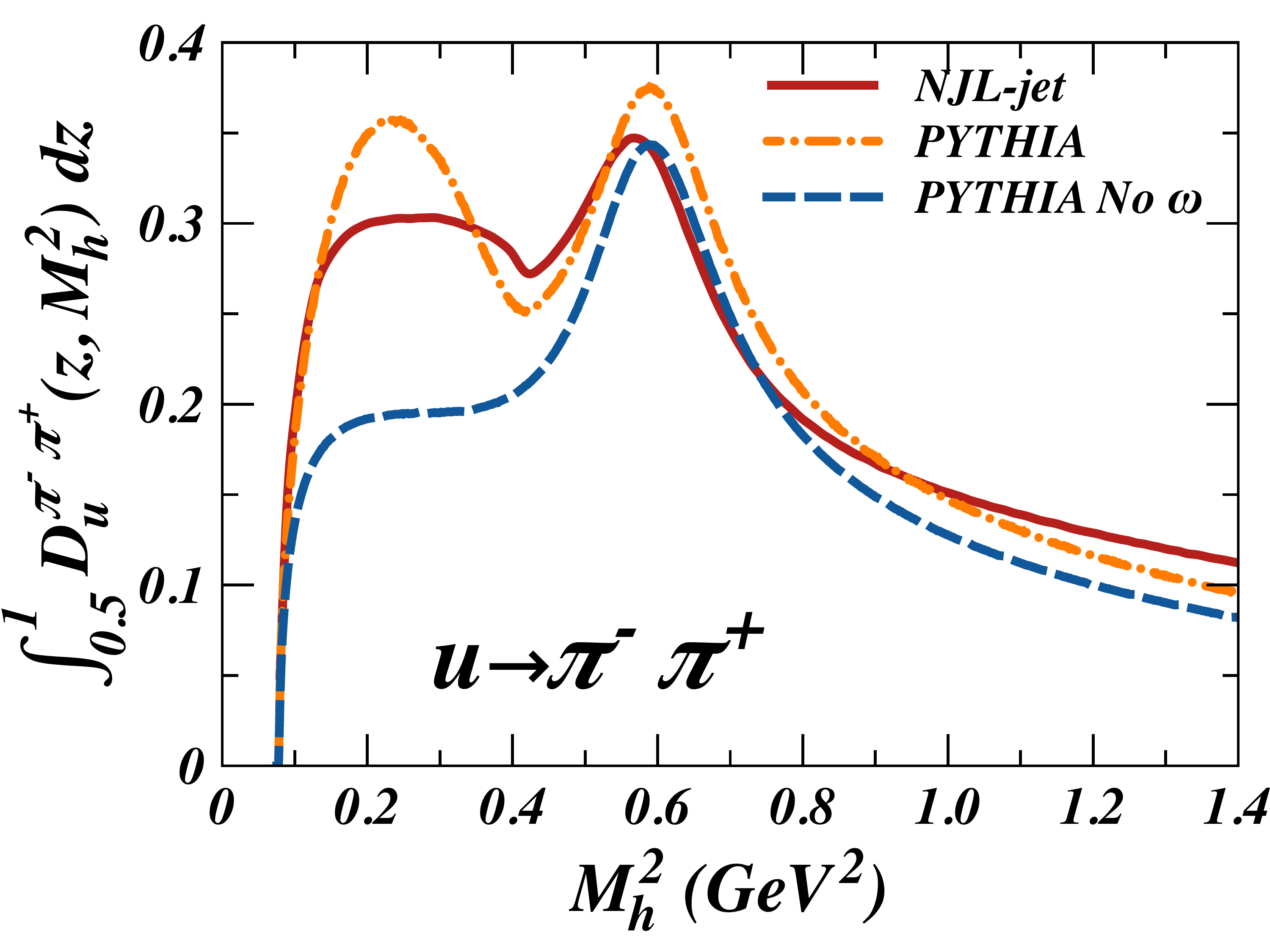}
}
\hspace{0.1cm} 
\subfigure[] {
\includegraphics[width=0.45\columnwidth]{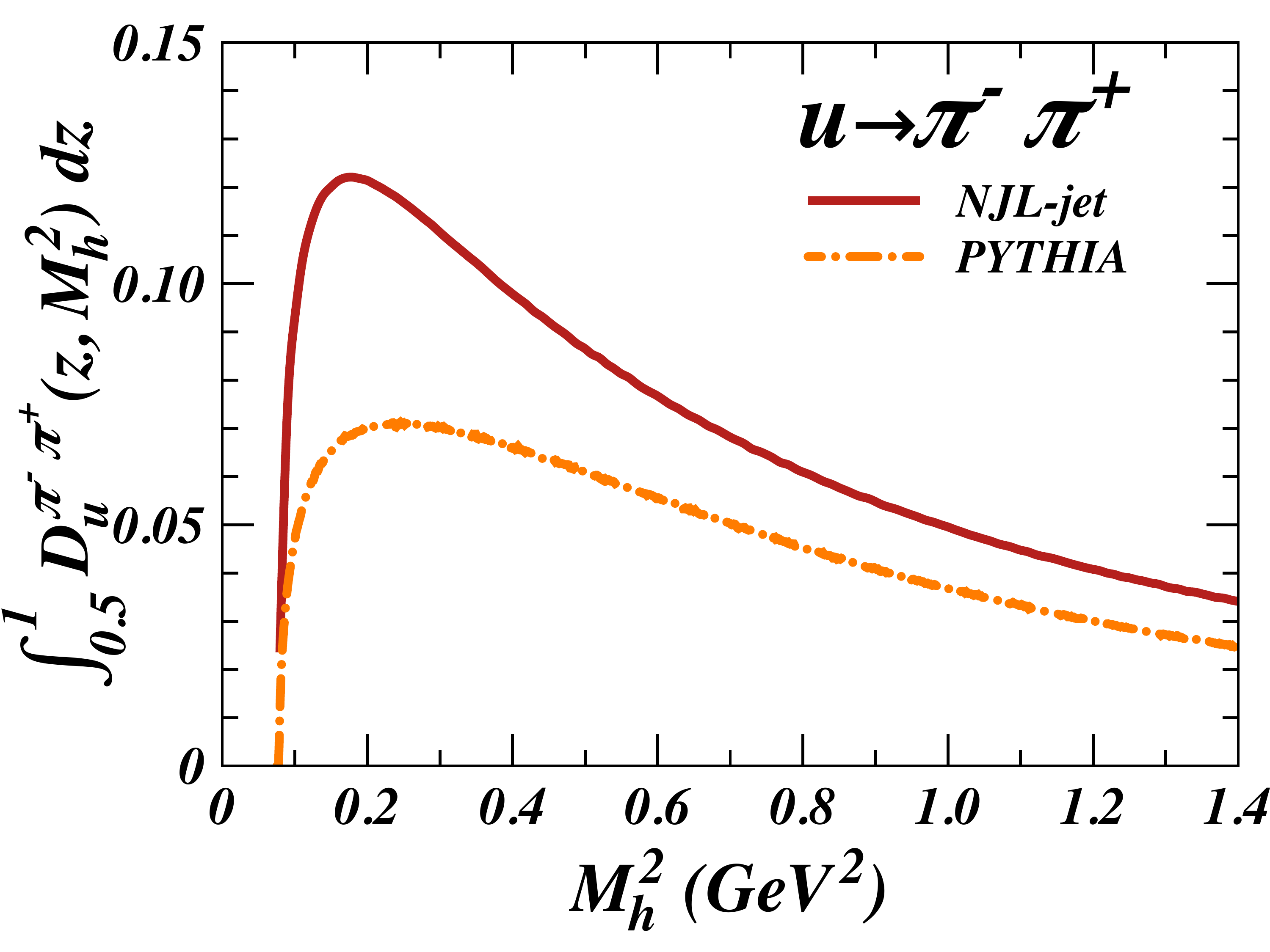}
}
\caption{The comparison of the results for $D_u^{\pi^+\pi^-}$ integrated over $z$, extracted from NJL-jet (solid red line) and PYTHIA (orange dot-dashed) simulations including (a) the primary and  (b) full final state pions. In plots (a) are also shown in blue dashed line the DFFs extracted from PYTHIA that exclude the $\omega$ meson decay products from the final state.  The MC simulations were performed with $8$ primary produced hadrons and over $10^{10}$ simulations.}
\label{PLOT_DFF_PYTH}
\end{figure}

  The results of the extraction of $D_u^{\pi^+\pi^-}$ from simulations of $u\bar{u}$ initial pair hadronization with the centre-of-mass energy $\sqrt{s}=20~\Ge$ are shown in Fig.~\ref{PLOT_DFF_PYTH}. The results, integrated over $z$ from $0.5$ to $1$, for (a) the full set of final hadrons and (b) only the primary produced pions depicted using orange dot-dashed line are compared to those from the NJL-jet model depicted using the red solid lines. Also shown using blue dashed line in Fig.~\ref{PLOT_DFF_PYTH}(a) is the DFF extracted using PYTHIA when we exclude the $\omega$ meson decay products from the final state hadrons. These results from PYTHIA clearly exhibit the same qualitative features as those from the NJL-jet model. The FFS results are much larger (up to $10$ times) than those including only the primary hadrons, though in PYHTIA the total probability of producing a $\rho$ meson is set to be equal to that of a $\pi$ meson of the same charge~\cite{Sjostrand:2006za}. Moreover, the shape of the invariant mass squared dependence of the FFS results clearly reflects the relevant vector meson spectrum giving the dominant contribution to DFFs. The $\rho^0$ meson peak and Breit-Wigner shape are seen in the range $0.4~\Gs$ to $0.8~\Gs$, while the $\omega$ meson contribution dominates the region from threshold to  $0.4~\Gs$. The last hypothesis is clearly confirmed when we examine the PYTHIA results that lack the $\omega$ meson decay products from the final state hadrons, where the large peak in the region below $0.4~\Gs$ is strongly suppressed along with a slight overall reduction of the DFF across the entire $M_h^2$ range. Similar effects are also observed when excluding $\omega$ meson decays in the NJL-jet model as well, though the results are not shown here to avoid overcrowding the plots.The results for the direct pions, when compared to the NJL-jet model results, reveal a very similar shape, but have different magnitudes because of the differences in the $z$ and transverse momentum dependences of the FFs between the two models and the vastly different scales.
  
  %===============================================================================
%===============================================================================

%%%%%%%%%%%%%%%%%%%%%%%%%%%%%%%%%%%%%%%%%%%%%%%%%%%%%
%%%%%%%%%%%%%%%%%%%%%%%%%%%%%%%%%%%%%%%%%%%%%%%%%%%%%
%%%%%%%%%%%%%%%%%%%%%   SECTION %%%%%%%%%%%%%%%%%%%%%%%%%%
\section{Conclusions and Outlook}
\label{SEC_CONC}

 In this work we have examined the effects of the vector meson decay products on the DFFs of pseudoscalar mesons on the example of $u\to\pi^+\pi^-$. We have considered two quark hadronization MC event generators: the MC framework based on the NJL-jet model and PYTHIA 8.1 based on the Lund string model. We extracted the DFFs from MC averages of the relevant two-hadron multiplicities, with the values discretized on a fine grid for the total light-cone momentum fraction and the invariant mass square of the hadron pair. DFFs were calculated using two sets of the final state hadrons: the primary hadrons directly emitted by the fragmenting quark and the full final state set that also includes the hadrons arising from the strong decays of the primary produced resonances (vector mesons in this case). Our analysis of both models has shown that the products of the vector mesons play predominant role in the contributions to DFFs for pseudoscalar hadron pairs, with the FFS results for $u\to\pi^+\pi^-$ DFF an order of magnitude larger than the results with only the direct pions. Moreover, the $M_h^2$ dependence of the $D_u^{\pi^+\pi^-}$ for FFS strongly resembles the vector meson spectrum relevant for the given mass range, with prominent $\rho^0$ peak and Breit-Wigner shape and $\omega$ peak shifted to the lower mass range due to the missing $\pi^0$. We note that for both models, the total probability of producing a $\rho$ meson is equal (PYTHIA) or lower (NJL-jet) than the probability of producing similarly charged pion.
 
  An intuitive explanation for these results can be given by simple combinatorial arguments within the NJL-jet model, for example considering $D_u^{\pi^+\pi^-}$. For simplicity, let's only consider the light quarks with $\pi$ and $\rho$ mesons as the only allowed hadronization channels. Also, let us limit the number of the primary produced hadrons to two. Then the only direct channel for a $u$ quark producing a $\pi^+\pi^-$ pair is
\al
{\nonumber
u\to d+\pi^+ \to u + \pi^+ + \pi^-,
}
while one of the secondary channels is
\al
{\nonumber
u\to d+\rho^+ \to u + \rho^+ + \rho^- \to u+ \pi^0 + \pi^+ + \pi^0 + \pi^-.
}
\begin{table}[h!]
\centering
\caption{The channels for producing the secondary $\pi^+\pi^-$ pairs in the $u$ quark fragmentation.}
\label{TABLE_SECONDARY}
\begin{tabular}{c|c|c} Primary & Final State & \# of $\pi^+\pi^-$ Pairs \\ \hline\hline
$ \pi^+ + \rho^-$ & $\pi^+ + \pi^0 + \pi^- $ & 1 \\
$ \pi^+ + \rho^0$ & $\pi^+ + \pi^+ + \pi^-$ & 2 \\
$ \pi^0 + \rho^0$ & $\pi^0 + \pi^+ + \pi^- $ & 1 \\
$ \rho^+ + \pi^-$ & $\pi^0 + \pi^+ + \pi^- $ & 1 \\
$ \rho^+ + \rho^-$ & $\pi^0 + \pi^+ + \pi^0 + \pi^-$ & 1 \\
$ \rho^+ + \rho^0$ & $\pi^0 + \pi^+ + \pi^+ + \pi^-$ & 2 \\
$ \rho^0 + \pi^+$ & $\pi^+ + \pi^- + \pi^+ $ & 2 \\
$ \rho^0 + \pi^0$ & $\pi^+ + \pi^- + \pi^0 $ & 1 \\
$ \rho^0 + \rho^+$ & $\pi^+ + \pi^- + \pi^0 + \pi^+$ & 2 \\
$ \rho^0 + \rho^0$ & $\pi^+ + \pi^- + \pi^+ + \pi^-$ & 4\\ \hline \hline
Total & & 17
\end{tabular}
\end{table}
 A summary of all the secondary channels is listed in Table~\ref{TABLE_SECONDARY}, which shows that the total number of $\pi^+\pi^-$ pairs from all the possible secondary channels is 17. Thus the DFFs in FFS are strongly enhanced by the wealth of available secondary hadron production channels, though one has to keep in mind that we have completely ignored the $z$ and transverse momentum dependence in this naive argument.

 Thus the main conclusion of this work is that in constructing models for  calculations of DFFs, a very careful and detailed description of fragmentation to and strong decays of the resonances in the relevant invariant mass region should be given. These details reflect strongly on both the shape and the magnitude of the DFFs.
 
  The NJL-jet model calculations of DFFs for both pion and kaon (as well as mixed) pairs will be presented in detail in our forthcoming article~\cite{Matevosyan:2013aa}. There, particular attention is given to presenting a realistic description of the vector meson two- and three-body strong decays, following the conclusions drawn in this work.
  
%%%%%%%%%%%%%%%%%%%%%%%%%%%%%%%%%%%%%%%%%%%%%%%%%%%%%
%%%%%%%%%%%%%%%%%%%%%%%%%%%%%%%%%%%%%%%%%%%%%%%%%%%%%
%%%%%%%%%%%%%%%%%%%%%   SECTION %%%%%%%%%%%%%%%%%%%%%%%%%%
\section*{Acknowledgements}

This work was supported by the Australian Research Council through Grants No. FL0992247 
(AWT), No. CE110001004 (CoEPP), and by the University of Adelaide. 

%%%%%%%%%%%%%%%%%%%%%%%%%%%%%%%%%%%%%%%%%%%%%%%%%%%%%
%%%%%%%%%%%%%%%%%%%%%%%%%%%%%%%%%%%%%%%%%%%%%%%%%%%%%
%%%%%%%%%%%%%%%%%%%%%   BIBLIOPGRAPHY %%%%%%%%%%%%%%%%%%%%%%
\bibliography{fragment}

\end{document}